# GuardianPWA: Enhancing Security Throughout the Progressive Web App Installation Lifecycle

Mengxiao Wang
*Texas A&M University*

Guofei Gu
*Texas A&M University*

***Abstract*—Progressive Web App (PWA) installation is critical for integrating web and mobile app functionalities, offering a seamless user experience. However, ensuring the security of the PWA installation lifecycle is essential for maintaining user trust and privacy. This paper introduces the GUARDIANPWA framework, a comprehensive approach to analyzing the PWA installation mechanism based on the CIA security principles (Confidentiality, Integrity, and Availability) and identifying areas where browser vendors fail to comply with these principles. Our study revealed 203 instances of non-compliance with security principles, highlighting how these irregularities in the PWA installation lifecycle can lead to potential violations of user privacy. For instance, in Firefox, PWAs installed in private mode incorrectly appear in normal mode, risking user confidentiality. Additionally, 29,465 PWAs are at risk because Samsung Internet does not display origins when PWAs navigate to third-party websites, undermining integrity. These findings were reported to browser vendors, leading to Firefox acknowledging four issues, resolving one, and planning to resolve two others. GUARDIANPWA supports developers by analyzing PWA manifest files for syntactic and semantic correctness, offering actionable recommendations, and helping to create PWAs that align with security best practices. By using GUARDIANPWA, developers and users can address critical security gaps and enhance compliance with CIA principles throughout the PWA installation lifecycle.**

## 1. Introduction

Progressive Web Apps (PWAs) represent a significant shift in web application development, combining the accessibility of traditional web pages with the functionality of native applications. Introduced in 2015, PWAs have steadily grown in popularity [1], [2], partly due to their unique features and adaptability on mobile devices [3]. Unlike conventional web applications, PWAs can be installed on mobile devices, offering an app-like experience while being lightweight and less resource-intensive [4]. They incorporate features such as offline capabilities, push notifications, and access to device hardware [5], which were traditionally limited to native applications. This seamless integration of web and mobile app characteristics has positioned PWAs as a versatile and efficient solution for both developers and users. The installation of PWAs is particularly important, as it ensures a native-like experience and improves user engagement by enabling background operations and offline access.

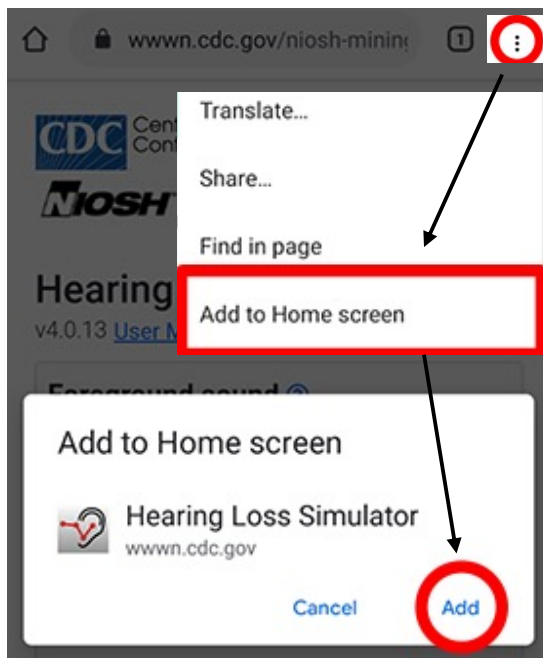

(a) Manual Installation

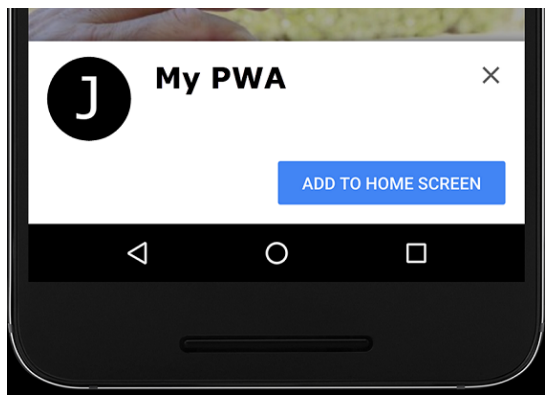

(b) Automated Installation

Figure 1: Installation Methods

PWAs can be easily added to a mobile device's home screen, providing a user-friendly, native app-like experience. Users can manually install PWAs from their browser, as shown in Figure 1a, or through an automated process triggered under specific conditions, illustrated in Figure 1b. Once installed, PWAs appear with their name and icon and operate full-screen without browser elements like the URL bar.

However, despite PWAs providing powerful features that can be permanently installed on user devices like Android apps, the PWA installation process does not introduce significant security considerations. From the perspective of browser vendors, these concerns can not escalate into severe issues such as Cross-Site Scripting (XSS) and therefore do not adhere to basic security principles when implementing the PWA installation mechanism. For example, browsers often fail to display the origin of the URL when navigating to third-party URLs, leaving users unaware of redirections and increasing the risk of credential leakage to untrusted websites. Moreover, inconsistencies in PWA installation mechanisms across browsers exacerbate these risks, leaving users with limited awareness and tools to mitigate potential vulnerabilities.

Our motivation stems from the need to analyze the PWA installation lifecycle through a structured security lens and evaluate its compliance with the foundational principles of security: Confidentiality, Integrity, and Availability (CIA). Currently, no framework or tool exists that enables a systematic investigation of the PWA installation mechanism. Moreover, PWAs behave inconsistently across different browsers, requiring us to create our own framework for analyzing PWA installations comprehensively. The challenges are further compounded by the fact that PWA installations can occur on both desktop and mobile devices, while browser versions are frequently updated. For example, Firefox version 127 used “Website” as the default name for PWAs without a specified name, making it difficult for users to distinguish between them. After we reported this issue, Firefox resolved it in version 128. These rapid browser updates highlight the need for extensive experiments to validate our findings and ensure reproducibility. The PWA installation process also involves significant user interaction, which cannot be fully addressed by automated tools. For instance, some browsers display the URL origin during installation, while others do not, necessitating manual verification. Conducting a systematic and comprehensive study of PWA installation security is therefore inherently challenging due to these complexities.

To the best of our knowledge, we are the first to analyze PWA installation security in a logically structured manner. We define the PWA installation lifecycle to facilitate a deeper understanding of the PWA installation mechanism and systematically identify violations of security principles. This lifecycle provides a foundation for addressing the unique challenges posed by PWA installation and allows for the identification of potential security issues. To address these challenges, we developed GUARDIANPWA, a framework designed to enhance security throughout the PWA installation lifecycle. It collects and aggregates PWA data to identify patterns, performs syntactic and semantic analysis during various phases of the lifecycle to uncover risks, and uses a semi-automated black-box mutation-based fuzzer (SABMFuzzer) during the pre-installation phase to efficiently discover anomalies that can violate security principles. Additionally, it combines manual verification with automated tools to detect abnormal behaviors and evaluate browser compliance with CIA principles. Finally, it proposes mitigation strategies, including reporting vulnerabilities to browser vendors and providing actionable tools for developers and users. Our contributions are summarized as follows:

- We analyze the PWA installation lifecycle using the CIA principles, identifying 203 cases of non-compliance that could lead to privacy violations or security risks.
- We introduce **GUARDIANPWA**, a comprehensive framework for evaluating and enhancing security in the PWA installation lifecycle. The framework supports developers by ensuring the syntactic and semantic correctness of PWA manifests and provides users with alerts about potential security threats.
- We report vulnerabilities to browser vendors and propose mitigation strategies, leading to real-world improvements in PWA security. For instance, Firefox has already resolved one of the identified violations and plans to address two more.

## 2. Background

This section overviews PWAs and their characteristics, the mobile installation process and its technical requirements, and the history, adoption, and real-world popularity of PWAs.

### 2.1. Characteristics and Installability of PWA

PWAs are web applications built using standard web technologies like HTML, CSS, and JavaScript, but with the functionality and feel of native apps. They offer various advantages such as push notifications, offline capabilities, and the ability to be installed on desktop and mobile devices.

PWAs can be added to the home screen from the browser with just a few steps, without the need for downloading from an app store. To make a PWA installable, it needs to (as documented [6]) meet certain technical requirements:

- **Web App Manifest:** A JSON file that defines the PWA’s appearance and behavior.
- **Secure Context:** The web app must be served over HTTPS or other secure local resources.
- **Service Worker:** Essential for offline functionality, push notifications, and caching.

However, based on our experiments, most browsers do not strictly enforce these requirements, which can lead to security concerns. We will discuss these concerns in detail in our repository [7].

### 2.2. PWA Ecosystem and Popularity

Despite early web-app efforts, after the 2007 iPhone native apps dominated thanks to better UX, faster loading, and direct hardware access such as cameras and microphones [8]. Web apps then lacked robust offline support and push notifications, and their icons were little more than glorified bookmarks.

Service Workers and Web App Manifests were the turning point, letting users install PWAs onto the home screen and integrate them with the OS [9]. Coined by Alex Russell in 2015 [9], PWAs gained vendor support soon after: in 2018 Google enabled mobile installation and Apple added Safari support, bridging web and native experiences [9].

PWA adoption is visible across official app stores [10]–[12] and third-party platforms [13]–[16]: the Google Play, Apple App, and Microsoft Stores all feature PWAs. Popular examples include Google Maps Go [17] (500M+ downloads), Twitter Lite [18] (50M+), and ScandiPWA [19]. Usage statistics reinforce this: Lyft PWA users take 11% more rides than native-app users, and 40% more users choose *Install PWA* over *Download App* [20], underscoring PWAs’ growing real-world significance.

## 3. Threat Model

We assume that the PWA has already been installed on the user's device, either being controlled by attackers or implemented by developers who may inadvertently make mistakes. Previous research has considered scenarios where attackers control a PWA and victims have granted sensitive permissions, such as notifications [21]. However, our focus is on the installation process itself, exploring the potential security risks arising during this critical phase.

In the case of malicious PWAs, the installation may occur through attacker-controlled mechanisms, such as misleading prompts, automatic browser prompts that users unknowingly accept, or overlapping UI elements that cause unintentional installation. On the other hand, benign PWAs developed by careless developers might still result in privacy violations due to basic yet critical mistakes. Our data analysis supports this assumption, revealing a frequent occurrence of such errors, which underscores the need for systematic examination of the installation process.

## 4. Methodology

### 4.1. Overview

Our methodology analyzes the PWA installation mechanism and its security risks in several steps. It begins with a preliminary study of two parts: analyzing the top 100 Tranco websites to identify common installation patterns, and reviewing all official PWA installation documents from the W3C and major browser vendors [22]–[28]. Additionally, we examined reported bugs concerning PWA installations to identify recurring issues [29]–[31].

From this preliminary study, we developed the PWA installation lifecycle, a comprehensive approach to identify key interactions and components throughout the PWA installation process. This lifecycle serves as the foundation for our research question: *What are the security principles violations associated with the current implementations of the PWA installation lifecycle, and how can they be mitigated?*

To address this question, we designed the GUARDIAN-PWA framework, a human-interactive approach with four components—Data Collection, Lifecycle Analysis, Practical Analysis, and Mitigation Strategies—that together identify, analyze, and address security-principle violations across the PWA installation lifecycle.

### 4.2. Preliminary Study

We conducted a preliminary study of the top 100 Tranco websites using Lighthouse and manual experiments for accuracy. We filtered out 39 inaccessible websites (either 404 errors or no longer available) and excluded them from our percentage calculations. Among the accessible websites, we found 45 without a manifest or with an invalid manifest. For example, *yandex.ru* has a manifest file containing only icons, and *blogspot.com* has a manifest with only the name and icons, lacking display fields. Of the accessible websites, 16 were installable PWAs. Among these, 10 did not use service workers, while 6 did.

Service workers are not required for installability but support offline and background functionality: over 62% of installable PWAs lacked one, showing the manifest is the key requirement (confirmed by manually installing them in Chrome). Chrome has since simplified its install criteria [32], though announcing this in a blog rather than official docs can confuse users and developers. We also discovered that the display mode impacts user experience. In the 16 installable PWAs, 9 used standalone mode and 7 used minimal-ui mode, with minimal-ui mode preferred for displaying the URL to users. Additionally, we manually reviewed websites with web manifests and found potential security risks. For example, *qq.com* includes a URL (*xw.qq.com*) whose `start_url` points to a different origin (*qq.com*). Although browsers mitigate this by using the current path, it underscores the need for systematic web app manifest analysis to prevent security risks.

TABLE 1: Analysis of the Top 100 Websites on Tranco

| Category | Number of Websites | Percentage |
|---|---|---|
| Inaccessible | 39 | 39.0% |
| Accessible | 61 | 61.0% |
| Invalid or no manifest | 45 | 73.8% |
| Installable PWA | 16 | 26.2% |
| Lack a SW | 10 | 62.5% |
| Have a SW | 6 | 37.5% |
| Tranco top 100 websites | 100 | 100% |

Table 2 summarizes PWA installation support across browsers, operating systems, and platforms; we use these supported browsers as the dataset for our installation, post-installation, and uninstallation analyses.

We reviewed official documents related to PWA installation from W3C and browser vendors [22]–[28], focusing on standards and guidelines. Key components of the Web Application Manifest include `start_url`, `display`, and `scope`, which are crucial for app launch, display mode, and URL scope, respectively. Misconfigurations can create security vulnerabilities, emphasizing the need for best practices [22].

We also analyzed reported bugs related to PWA installations to identify recurring issues across different browsers. Our review revealed issues such as off-scope navigation and discrepancies in bug responses. Of the CVE bugs related to PWA installation [33], only two of seven were relevant. The first [34], [35] involved overlaying PWA prompts onto other origins, and the second [36], [37] allowed arbitrary app installations through the service worker scope. These have been fixed by vendors. From 38 Webkit bugs [29], only one related to PWA installation [38], where PWAs open external links but stay in standalone view. We discovered that this issue remains unfixed in the Samsung Internet Browser. Out of 15 Firefox bugs, five were related and valid for PWA installation [31]. Chromium-based browsers had several vulnerabilities, including background tabs launching

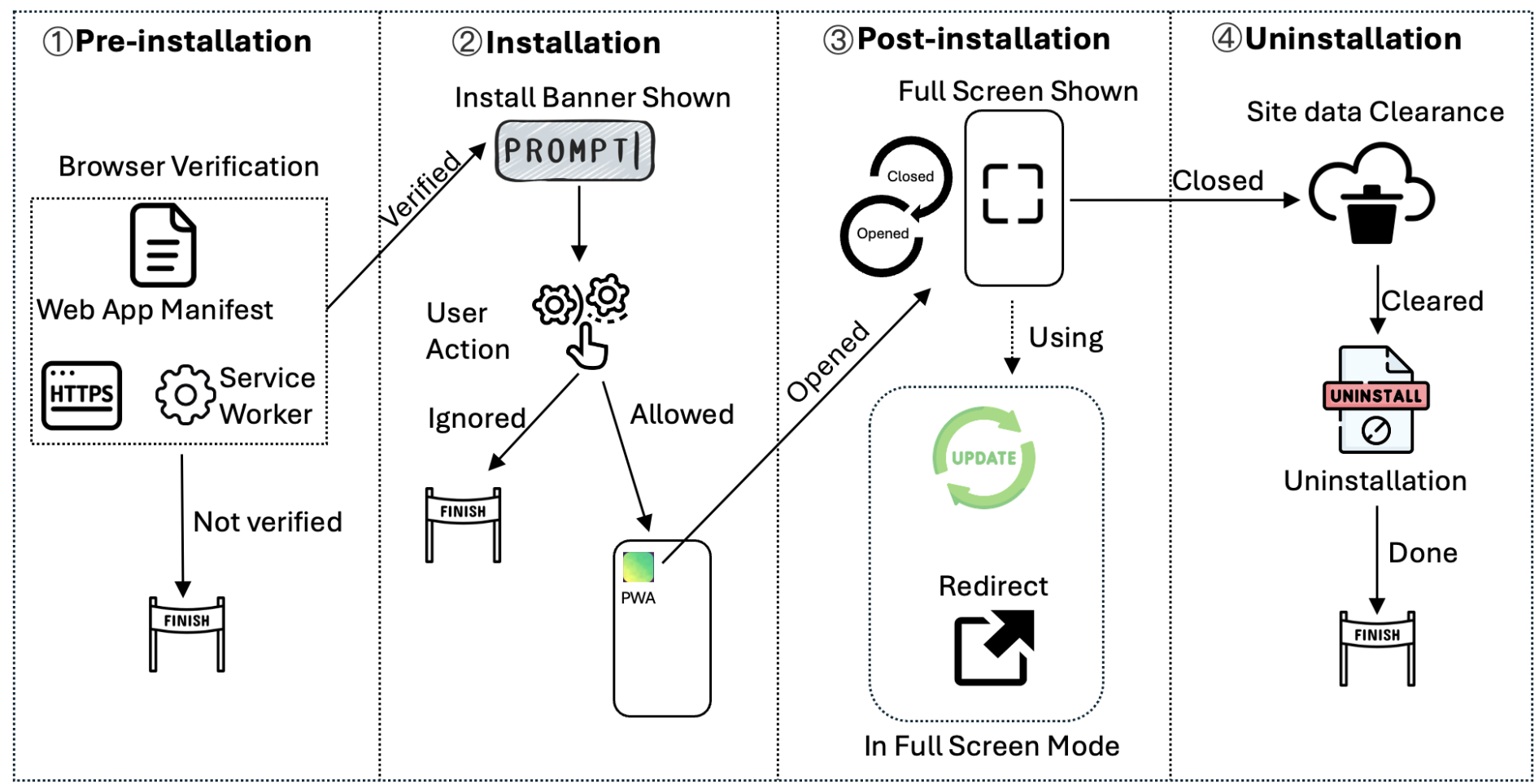

Figure 2: PWA Installation Lifecycle

TABLE 2: Browser Support Across Different OS and Platform

| Platform | OS | Safari | Firefox | Chrome | Edge | Opera | Brave | Samsung Internet | Tor Browser |
|---|---|---|---|---|---|---|---|---|---|
| Desktop | Linux | † | ○ | ● | ● | ○ | ○ | † | ○ |
| | macOS | ◐ | ○ | ● | ● | ○ | ○ | † | ○ |
| | Windows | † | ○ | ● | ● | ○ | ○ | † | ○ |
| Mobile | iOS | ● | ● | ● | ● | ○ | ○ | † | ○ |
| | Android | † | ● | ● | ● | ● | ● | ● | ○ |

●: PWA installation is supported, ○: PWA installation is not supported, †: Browser is not supported in this OS or platform, ◐: PWA can be installed by *Add to Docker*

PWAs without user interaction [39], dynamic PWA icons changing post-installation [40], repeated PWA installations from pressing *Enter* [41], and persistent installation dialogs across origins [42], [43]. Other issues involved hidden permission prompts in standalone mode [44], tapjacking vulnerabilities in Android Chrome [45], installations from sandboxed pages [46], and overlaying install prompts [47]. These have been addressed by browser vendors.

Existing bug reports focus on the installation and post-installation phases while neglecting pre-installation and uninstallation, motivating our lifecycle analysis of security-principle violations and risks.

## 5. PWA Installation Lifecycle Outline

Based on the review of the official documentation and preliminary results, we design the PWA installation lifecycle in Figure 2, which consists of four main phases: Pre-installation, Installation, Post-installation, and Uninstallation.

In *Pre-installation*, the browser verifies the PWA's requirements (a valid manifest, service worker, and HTTPS). If met, *Installation* shows a prompt the user can accept or ignore; accepting installs the PWA and adds an icon. In *Post-installation*, the PWA runs full-screen with a native-like experience and can update automatically. Finally, *Uninstallation* removes the app and clears its site data. Throughout, the browser handles verification, prompting, updating, and data clearance, while the user decides on installation and removal.

## 6. Analysis

Our analysis focuses on the PWA installation lifecycle, specifically the four phases introduced in Section 5. For each phase, we will conduct a practical analysis of potential security risks using our crawled PWAs and affected browsers. We will discuss the possible consequences of these risks in detail, providing a thorough understanding of the security implications during the PWA installation process.

### 6.1. Overview of the Security Principles Violations

Our analysis identified 203 security-principle violations across PWAs. These risks are prevalent across various browsers and represent the most critical vulnerabilities that are currently prioritized for future fixes and enhancements. We outline each identified risk's potential impact on users and specify which browsers are affected. Rather than enumerate all violations here, we provide the complete list in Appendix A.4, with technical descriptions and demonstrations in our repository [7]. To address these security concerns effectively, we will discuss detailed mitigation strategies in Section 7.

### 6.2. Pre-installation Phase Analysis

In this subsection, we discuss the analysis conducted during the pre-installation phase of PWAs. This includes examining user browser environments and installation criteria.

**6.2.1. Inconsistencies in PWA Profile and Mode Isolation.** According to Table 3, PWA multiple-profile support is inconsistent across browsers and platforms. On desktop (Linux, macOS, Windows), Chrome and Edge support multiple profiles and isolate them, so each profile keeps separate cookies and permissions.

On Android, however, Firefox, Chrome, and Edge support multiple profiles but do not isolate them: permissions such as camera, microphone, and geolocation are shared across profiles and can be reused without the user's knowledge. Brave and Samsung Internet support no profiles at all, forcing a single set of permissions and cookies for all activity. On iOS, by contrast, most major browsers (Safari, Firefox, Chrome, Edge) support multiple profiles with proper isolation—a more consistent and secure experience than Android.

Based on our experiments, only Firefox on Android supports installing a PWA from private mode. Table 4 contrasts user perception with actual behavior: users assume that cookies, cached files, browsing history, and form input are discarded after a private session, but these data persist exactly as in normal mode. Cookies, for instance, are still retained, so tracking continues just as in normal mode—undermining the privacy users expect from private browsing.

Such cross-platform inconsistency confuses users, who expect uniform isolation across devices but instead face behavior that varies widely between desktop and mobile.

**6.2.2. Discrepancies and Flaws in PWA Installation Requirements.** We analyze manifest fields both syntactically and semantically. Although most fields are optional, some are required for installability (e.g., fullscreen mode) and automatic installation. Syntactically, many PWAs and browser vendors ignore the W3C's prescribed formats, which can make PWAs non-installable; semantically, we run the crawled PWAs through browsers to surface the resulting security concerns.

We developed a **semi-automated black-box mutation-based fuzzer (SABMFuzzer)** that combines syntactic analysis guided by official documentation with semantic fuzzing of abnormal manifest values. It focuses on the fields that most affect user consent and awareness—`Name`, `Short Name`, `Start URL`, `Display Mode`, and `Icons`: misconfigured names cause confusion, abnormal URLs redirect to malicious sites, wrong display modes hide critical information, and misleading icons misrepresent the PWA.

We manually defined the manifest schema from official documentation and designed algorithms to mutate field values systematically, covering all documented values plus undocumented fields to assess their effect on installation and functionality. We also adapted common web attacks such as XSS to test whether injected scripts or altered URL parameters could exploit a PWA.

The process is semi-automated: a script mutates manifest fields periodically (e.g., every minute), while installation and behavioral evaluation are done manually to catch subtle issues such as delayed field updates or unpredictable installation behavior. We repeated each experiment three times to reduce missed anomalies. Combining fuzzing, manual verification, and repetition, SABMFuzzer reliably uncovers manifest vulnerabilities and yields actionable security insights.

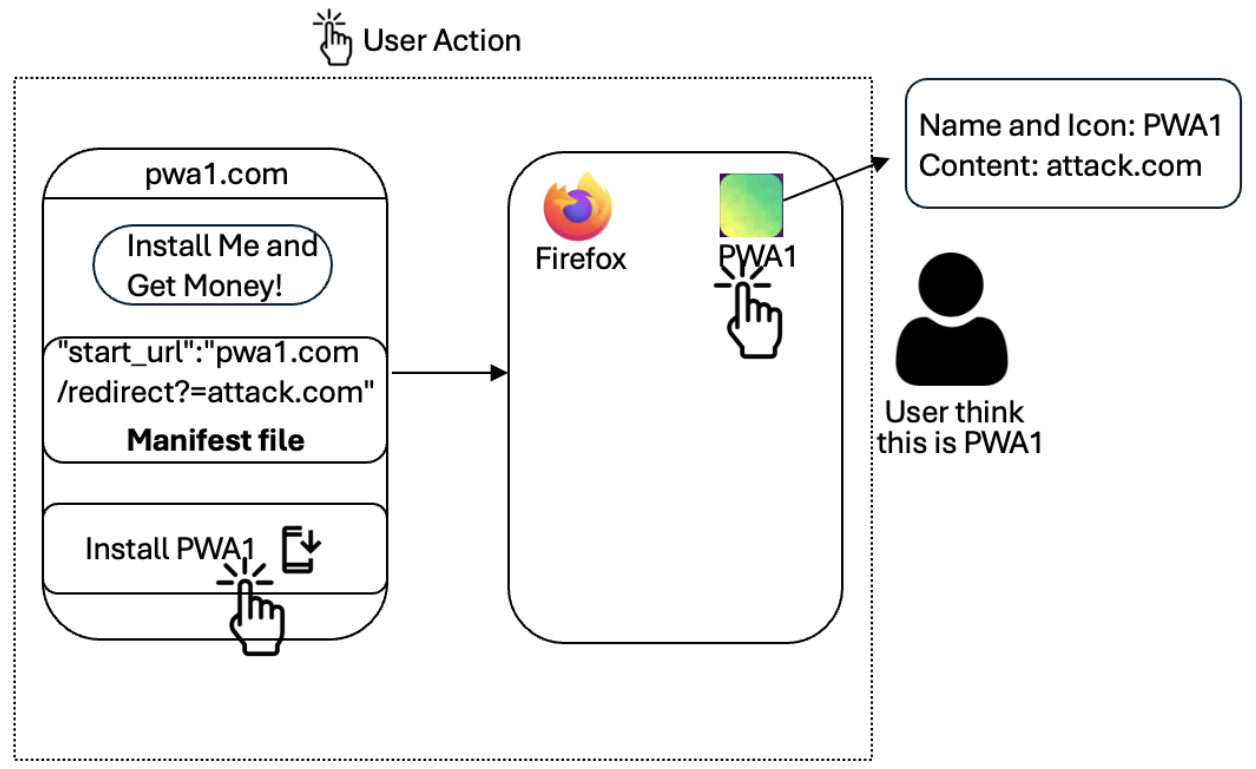

Figure 3: Craft Start_url leads to DeceptiveInstall

**Misuse of Start_url and Scope.** The *start_url* field specifies the page loaded when the app starts. The specification rejects URLs that violate standard syntax and forbids `https://` URLs unless they match the PWA's own domain, preventing cross-domain requests. Yet our fuzz testing shows relative URLs are still accepted: as in Figure 3, an attacker can use relative paths or query parameters (e.g., *?redirect=attack.com*) to redirect users to malicious sites while they believe they remain within the trusted PWA domain—enabling phishing that tricks users into installing software or entering credentials.

This underscores the need for robust *start_url* validation that keeps navigation within the PWA's trusted domain. In our dataset of 277,667 PWAs with a *start_url*, many use relative paths such as *../* or *../../*, and 233 use tracking-style parameters like *?id* that can build personalized paths to track users—a semantically poor practice.

Additionally, 226 PWAs use HTTPS but are not on the same origin as the PWA, which, while ignored and defaulted to the current path, is syntactically incorrect. Furthermore, 198 PWAs use parent paths, posing security risks as users may see and install different PWAs from different origins. Finally, 27 PWAs have an empty *start_url* value, which is also syntactically incorrect.

The *scope* field can bypass the Same-Origin Policy, letting an installed PWA redirect to other sites. As with *start_url*, many PWAs use relative or parent paths for *scope*, so users may install a PWA different from what they perceive. Of the 214,694 PWAs using *scope*, 33 use parent paths (semantically incorrect and a privacy risk), 75 use HTTPS

TABLE 3: PWA Multiple Profile Support Across Different Browsers and Platforms

| Platform | OS | Safari | Firefox | Chrome | Edge | Opera | Brave | Samsung Internet | Tor Browser |
|---|---|---|---|---|---|---|---|---|---|
| Desktop | Linux | † | ○ | ● | ● | ○ | ○ | † | ○ |
| | macOS | ● | ○ | ● | ● | ○ | ○ | † | ○ |
| | Windows | † | ○ | ● | ● | ○ | ○ | † | ○ |
| Mobile | iOS | ● | ● | ● | ● | ○ | ○ | † | ○ |
| | Android | † | ◐ | ◐ | ◐ | ◐ | ◑ | ◑ | ○ |

●: Browser supports multiple browser profiles for PWA installation and isolates these PWAs, ○: PWA installation is not supported, †: Browser is not supported in this OS or platform, ◐: Browser supports multiple browser profiles for PWA installation and does not isolate these PWAs, ◑: Browser does not support multiple profiles

| Characteristic | User Perception | Actual Behavior |
|---|---|---|
| **Cookies** | Non-persistent | Persistent |
| **Temporary File Cache** | Non-persistent | Persistent |
| **Browsing History** | Non-persistent | Persistent |
| **Form Input Data** | Non-persistent | Persistent |

TABLE 4: Misunderstandings of User Perception and Actual Behavior for PWA Installed from Private Mode

on a different origin, and 18 are empty—syntactic errors that can break installation.

**Misuse of Name and Icon.** The *name* field specifies the name of the application. When it exceeds 1000 characters, desktop browsers (e.g., Chrome, Edge) will not install the PWA, deeming it abnormal activity. However, mobile devices will still install it. As it is plain text, there is no risk of script or code execution or command injection.

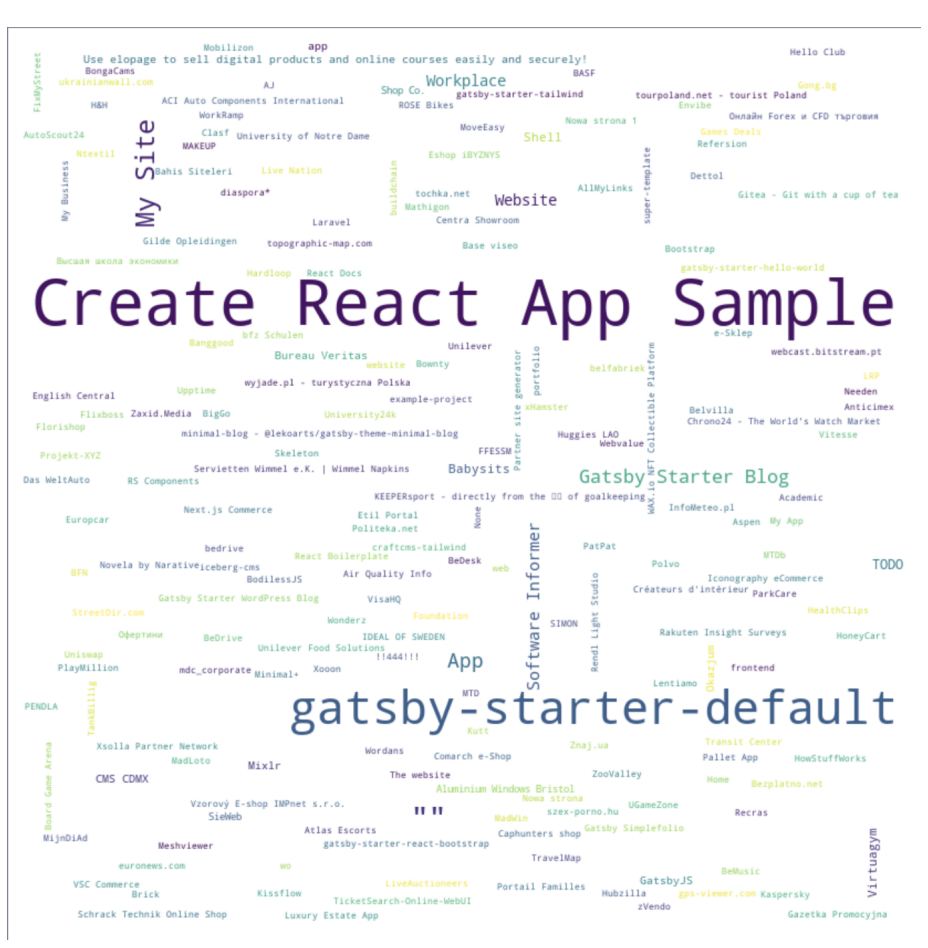

Figure 4: Word Cloud for name field

Many browsers assign a default name to PWAs if developers do not specify one or leave it empty. In our experiments, on desktop browsers, a PWA without a name has no install button. On iOS devices, users can define their names for PWAs. On Android, except for Firefox, all other browsers either use the URL or a user-defined name. Firefox, however, uses *website* as the PWA name, which can cause integrity issues if different domains do not explicitly specify a name.

The use of duplicate names for PWAs poses significant security risks, as it can confuse users and increase the chances of phishing attacks. Figure 5 illustrates this issue with four PWAs that have the same name and icon but come from different domains and contain different content. This similarity makes it difficult for users to identify the legitimate PWA, thus increasing the risk of falling for fraudulent applications. Our data supports this concern. Figure 4 shows that many PWAs unintentionally share the same name, often because developers use templates without renaming them. This leads to both syntactic and semantic errors. For example, "Create React App Sample" is the most commonly duplicated name, used by 6456 PWAs. Additionally, 838 PWAs have empty names, and there are 5525 unique duplicate names, affecting a total of 30634 PWAs.

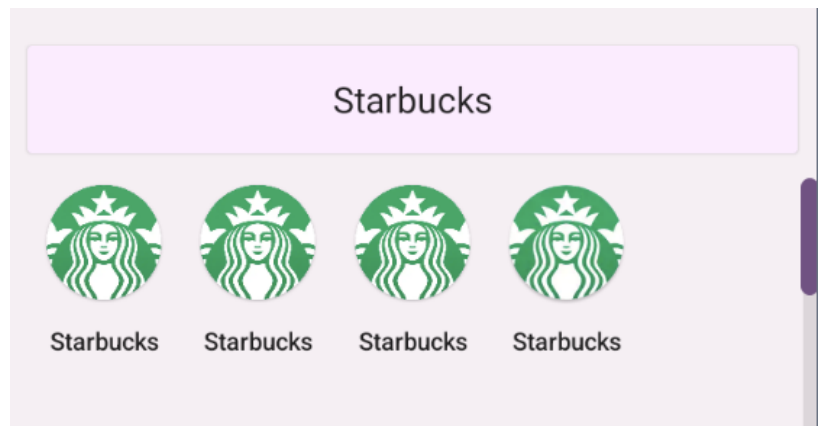

Figure 5: Same name and icon for Starbucks

**History Sniffing via Chrome Mini-Infobar.** A local attacker is defined as an adversary who gains control of the user's machine at a specific time, $T$. This attacker should not be able to access any historical information about user actions taken before $T$. Gaurav et al. [48] defined a local attacker in a similar way, and we adopt the same definition for our threat model.

In our investigation, we discovered that Chrome version 120.0.60 contains a vulnerability involving the mini-infobar prompt for PWAs. Specifically, this mini-infobar is triggered for PWAs meeting specific criteria, even when the user has cleared their browsing history. An attacker can systematically visit PWAs that trigger the mini-infobar (approximately 200,000 tested PWAs initially met the criteria). By observing whether the mini-infobar appears or not, the attacker can infer whether the user has interacted with specific PWAs in the past. The root cause of this issue lies in a feature introduced in older versions of Chrome. If a user ignored the mini-infobar prompt for a specific PWA, Chrome would optimize the user experience by not prompting the mini-infobar for that PWA again in the future. While intended as a usability enhancement, this feature inadvertently enabled attackers to infer a user's browsing history based on the absence or presence of the mini-infobar.

This issue was fixed in subsequent versions of Chrome, where stricter mini-infobar prompt rules were implemented. Additionally, Chrome no longer records whether a mini-infobar prompt is shown, regardless of whether the user has previously interacted with the PWA. Although Chrome resolved the vulnerability following our disclosure, users of older Chrome versions remain susceptible to this issue. This vulnerability allows an attacker to infer a user's browsing history even after the history has been cleared, thereby violating the confidentiality principle of the CIA security triad.

## 6.3. Installation Phase Analysis

**6.3.1. Abuse of Installation Banners for Phishing Attacks.** The installation banner introduces several risks: overlapping prompts cause unintended clicks, missing confirmation enables accidental installs, hidden origins enable phishing, automatic prompts mislead users, and allowing duplicate PWAs lets users install PWA2 while believing it is PWA1, risking malicious redirects.

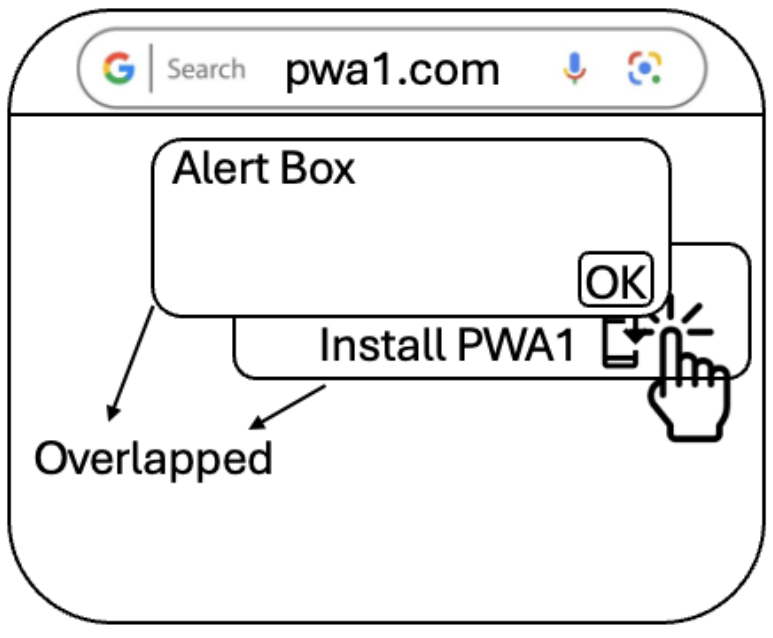

Figure 6: Alert prompt overlap with PWA installation banner

**Dialog Boxes Overlap with PWA Installation Banner.** On desktop Chrome and Edge, alert/confirm/prompt dialogs take priority over the install prompt. A user clicking too quickly can inadvertently hit the install prompt and install the PWA unawares (Figure 6); unlike mobile, desktop offers no secondary confirmation, enabling unintended installs. Moreover, an alert box can accept a user-typed origin URL, so an accidentally installed PWA may auto-redirect to that site.

**Lack of Banner Confirmation.** Without a confirmation step—common in Android Chrome and desktop browsers—users can unknowingly install a harmful PWA. Automatic install prompts worsen this in Android Chrome and Edge, tricking users into installing a PWA that looks legitimate.

**Obscured Origins and Problematic Multiple Installations.** Hidden origins and duplicate installs both create risk. In Firefox on Android, the full-screen install banner does not show the PWA's origin (Figure 7), easing deception.

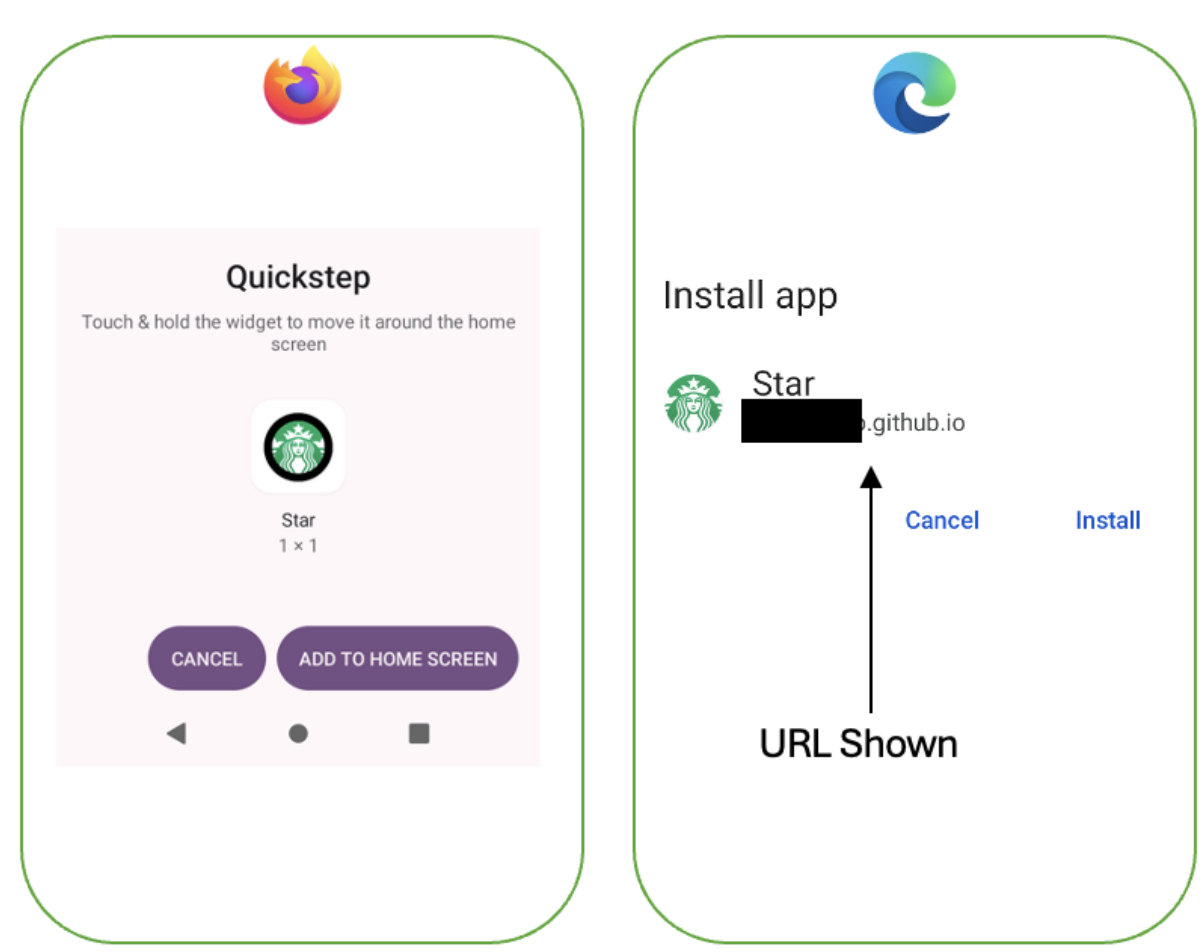

Figure 7: Origin Visibility Comparison between Firefox and Edge on Android

Allowing multiple identical PWAs further enlarges the attack surface; most browsers permit this, except desktop browsers, Android Chrome, and Brave, which block reinstalling an already-installed PWA—good practice, since it stops attackers from cloning a PWA's name and icon.

## 6.4. Post-installation Phase Analysis

**6.4.1. Chaotic and Inconsistent PWA Update System.**
**Delayed Manifest Update.** Only desktop Chrome/Edge and Android Chrome detect manifest updates, and only every 30 days (daily once a change is seen) [49], [50], so manifests can stay outdated. Only the name, short_name, display, start_url, theme_color, and scope fields trigger updates; even then, sensitive fields like display, start_url, and scope are not changed in the existing PWA, while name is. Android Chrome and desktop Edge notify users of a name change but do not apply it, whereas desktop Chrome asks consent to update—an inconsistency that confuses users and can enable phishing.

**PWA Update Stalled by SW Cache.** Service workers can adopt a *Cache Only* strategy that serves content solely from the cache, so some PWAs never update automatically and require manual cache deletion. Because the cached code is not refreshed, users may unknowingly run outdated, vulnerable scripts that attackers can exploit (e.g., known vulnerabilities or phishing content), making cache-strategy management important to PWA security.

**6.4.2. Navigate to 3rd-Party Without Showing Origin.** Navigating to third-party scopes poses notable risks. In a normal browser, a redirect to a third-party URL is visible in the address bar, but a full-screen PWA hides the URL bar, so users cannot see the site they are on. Most browsers now show the URL in full-screen mode, but Samsung Internet still does not. We have identified five primary scenarios that can lead to such a navigation issue:

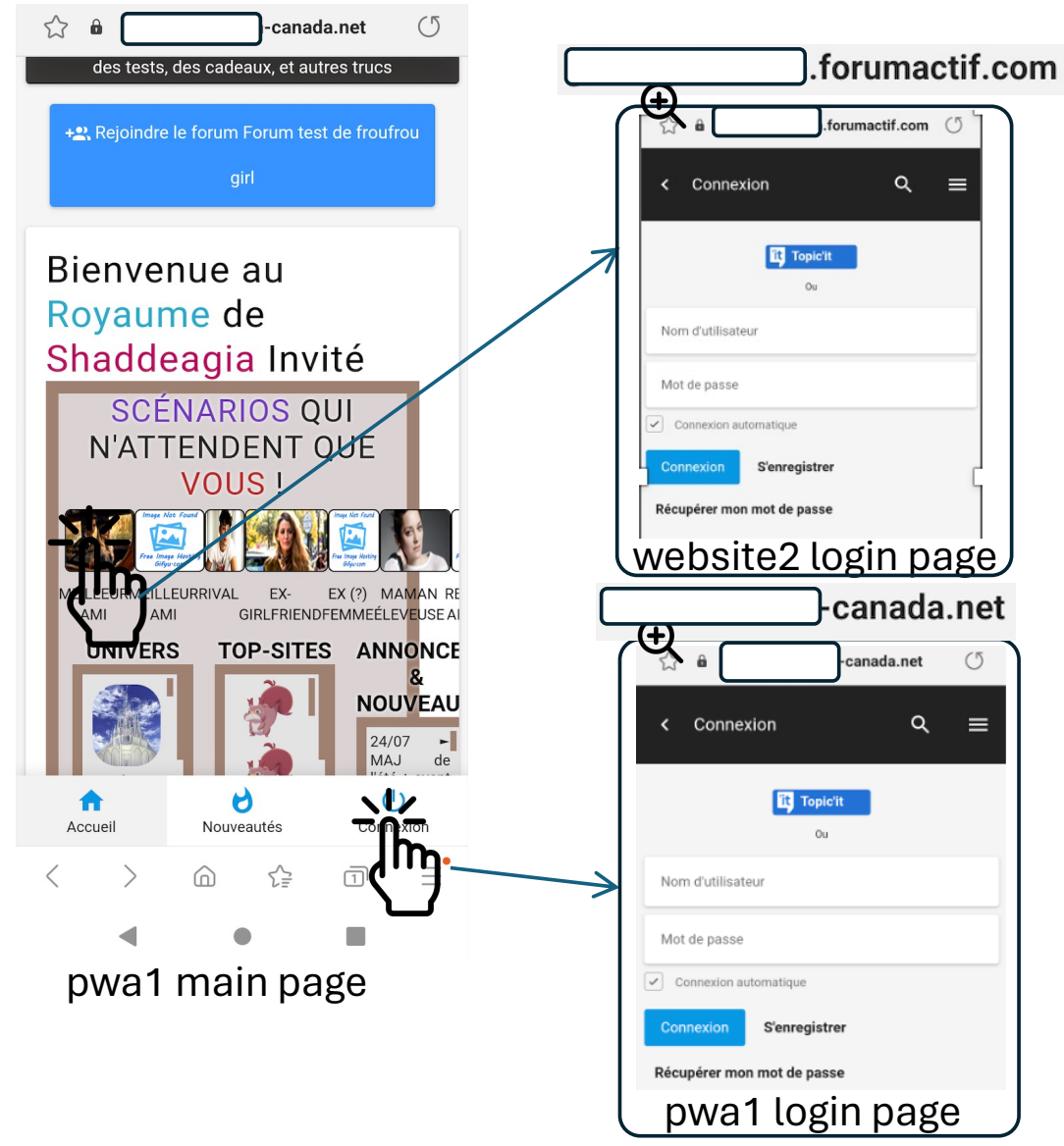

Figure 8: Navigate 3rd-Party Scope real-world example

- **URL Parameter Redirect + start_url Manifest**: When a PWA is launched with a URL parameter that redirects to a third-party site from the start_url specified in the manifest, users can not notice the change in scope.
- **Meta Refresh Redirect**: A meta refresh tag can automatically redirect users to a different URL after a specified time interval, potentially leading to third-party sites without user awareness.
- **JavaScript Redirect**: Various JavaScript methods can redirect users to new URLs, such as *window.location.assign*, *window.location*, and *window.location.href*. All these methods can seamlessly redirect users within a PWA.
- **HTML Link Redirect**: Clicking on a hidden or obfuscated HTML link (<a>) can redirect users to a third-party site without their knowledge.
- **CSS/HTML Redirect**: Techniques involving hidden links or other HTML/CSS manipulations can redirect users to third-party sites without noticeable UI changes.

If such redirections occur without the user's knowledge, several serious consequences can follow:

- **Phishing, Data Theft, and Unauthorized Actions**: Users tend to trust a redirected site, so they may hand login credentials or payment details to a malicious replica—e.g., a fake bank login page that captures their credentials and grants attackers full account access.
- **Iframe Full-Screen Exploitation**: Among the top 10,000 Tranco sites, 52.5% (3768 of 7178 tested) allowed full-screen iframes because of missing or permissive *X-Frame-Options* headers. Attackers can embed these as realistic replicas of trusted sites, deceiving users into sharing data under a false sense of legitimacy.

In Figure 8, two visually identical pages have different URLs, which can trick users into entering credentials on a malicious site. Our analysis found 29,465 PWAs vulnerable to this attack. Because users cannot tell they are on a third-party site without a visible URL or UI change, browsers must clearly signal such transitions to prevent phishing.

## 6.5. Uninstallation Phase Analysis

**6.5.1. Concealed and Inconsistent Clear Site Settings.** Clearing site settings is central to privacy. For native apps, deleting the app removes all associated settings, permissions, cookies, and cached files, or clears them on reinstallation. PWAs differ: clearing a PWA's site settings may not fully erase them, and a PWA can keep sending notification requests even after uninstallation. Users assume uninstalling resets all permissions, but it often does not—opening the door to phishing. Examining clearance granularity, we found that some permissions and cached data persist after a manual clear, risking privacy.

**6.5.2. Opaque and Residual Uninstallation.** Uninstalling a PWA is less intuitive than removing a native app. On Android Chrome, PWAs are WebAPKs offering *Remove* (deletes only the icon) and *Uninstall* (removes the PWA); neither clears site settings or history. On desktop, dragging a PWA to the trash does not uninstall it—users must use the PWA's own interface—so many leave it installed, still able to send notifications usable for phishing. Because PWAs do not appear in OS application lists, they are hard to find and remove, and users unknowingly retain residual data and permissions that increase their exposure to unwanted notifications and phishing.

**Persistent Notifications after Uninstalling.** Even after uninstallation, a PWA can still send notifications. Chrome and Samsung Internet show only the PWA's name, not its URL, and the name can change at any time, so users are easily deceived into clicking malicious notifications—an effective phishing vector.

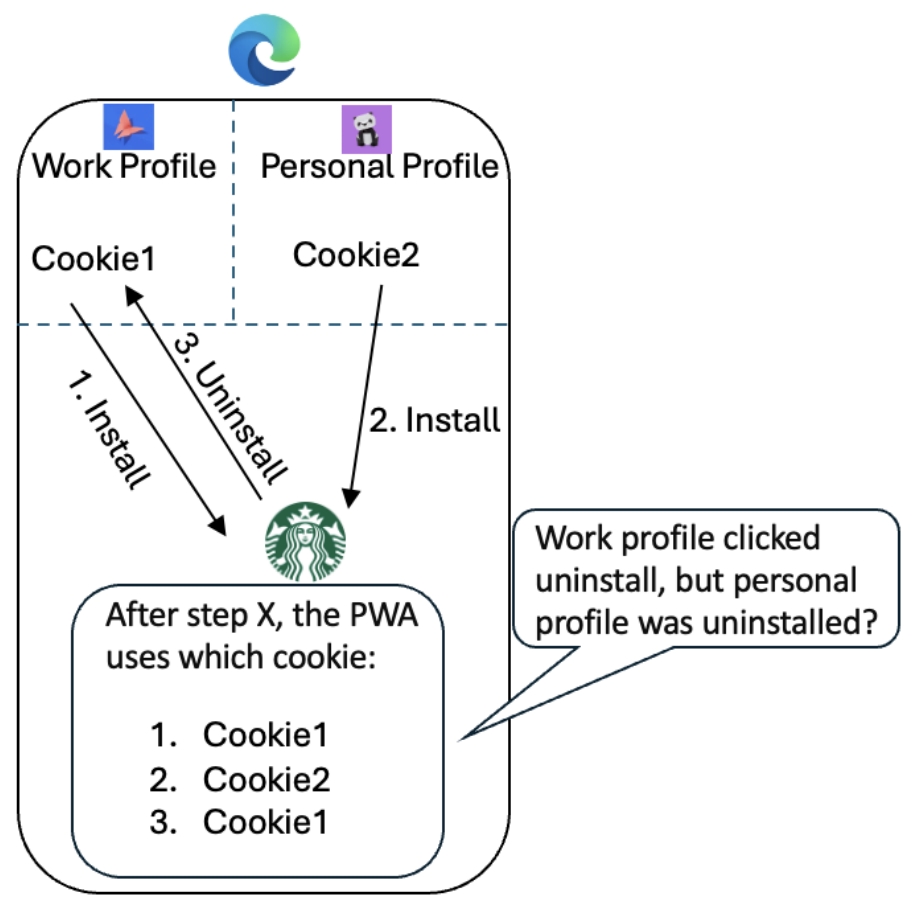

Figure 9: Mismatch in Uninstallation Across Different Profiles

**Different Profiles Managing the Same PWA.** Chrome groups installed PWAs under a single entity across profiles, so uninstalling from one profile can inadvertently affect another (Figure 9); identical profile avatars make this harder to notice. Edge instead forbids profile switching for PWAs: the profile that installed a PWA must uninstall it (via *edge://apps/all*) before another can manage it. Both behaviors risk residual data, unauthorized PWA access, and phishing, so users must understand profile-specific management.

TABLE 5: Risk Breakdown by CIA Dimensions

| **CIA Dimension** | **Risk Count** |
|---|---|
| C (Confidentiality) | 84 |
| I (Integrity) | 114 |
| A (Availability) | 5 |
| **Total** | **203** |

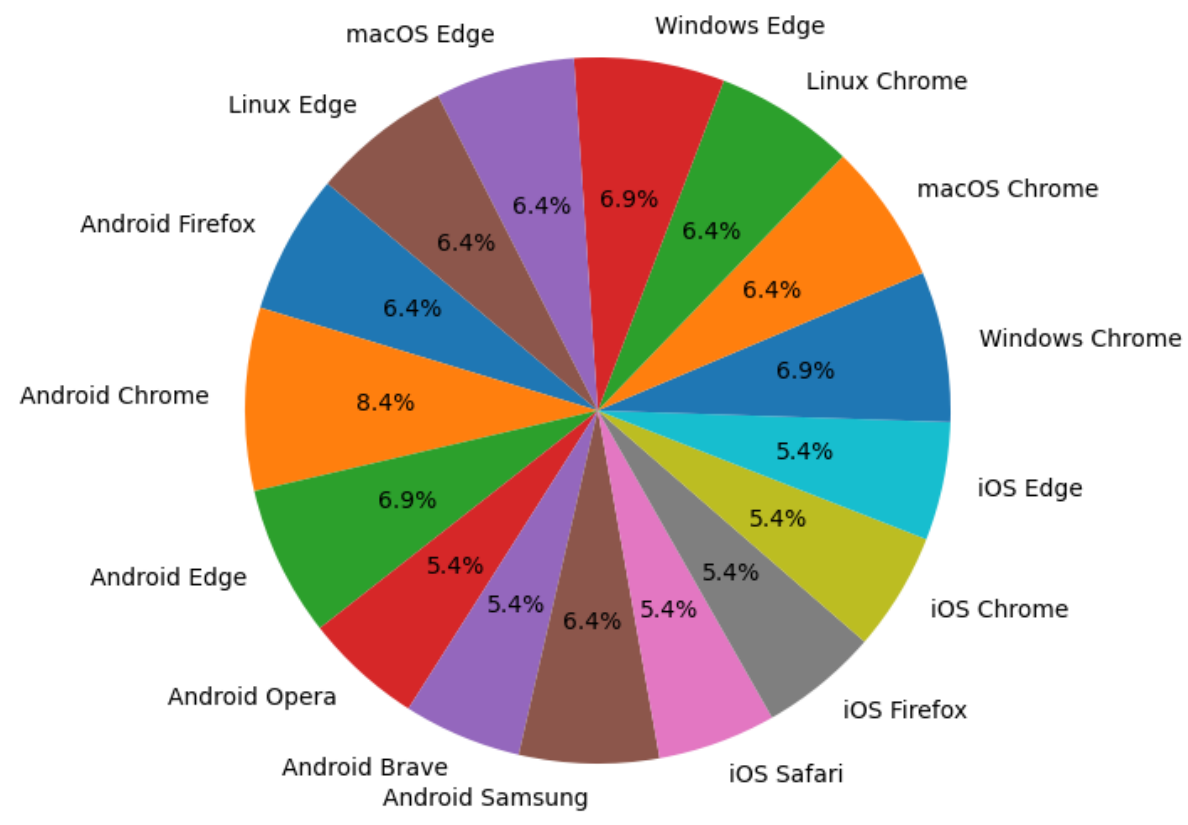

Figure 10: Security Principles Violations per "Operating System + Browser"

## 6.6. Overview of Security Principle Violations

Our systematic study of the PWA installation lifecycle reveals that although browsers are meant to serve as a trusted interface between users and web applications, they often fail to uphold this guarantee. Figure 10 illustrates the distribution of security violations across the sixteen "OS + Browser" combinations. Notably, the most popular combination, *Android Chrome*, exhibits more violations because it supports numerous additional PWA installation features compared to other browsers. This, however, does not imply that fewer features automatically enhance security. For instance, *Android Samsung Internet* implements fewer PWA features yet still incurs a relatively high number of violations. Moreover, Table 5 indicates that most violations concern *integrity* and *confidentiality*, which carry significant implications for user privacy and trust. These findings highlight the need for more consistent and secure implementation practices across different platforms and browsers—especially in light of PWA functionalities that blur the boundaries between native and web applications.

# 7. Mitigation Strategies

## 7.1. Overview

No browser is free of issues, though some are more secure than others; Chrome offers the best user experience. With proper security practices, users can still use these browsers safely.

Identifying problematic PWAs in the pre-installation phase can prevent their installation, but this needs developers, browsers, and users together. Stricter browser enforcement would spare non-expert users extra tools; where browsers fall short, developers can use GUARDIANPWA-Developer to catch the security flaws our analysis shows many introduce unintentionally. If developers still introduce risks, users can rely on GUARDIANPWA-User, which assesses a PWA in the pre-installation phase so risks surface before installation. For an already-installed PWA, we provide accurate uninstallation steps, or the simplest fallback: reinstalling the browser to purge all PWA data.

## 7.2. Browser Perspective

Browsers should strictly adhere to security principles; doing so would spare non-expert users from needing additional tools.

For each phase of discovered security risks, we evaluate whether to report them to browser vendors. If a security risk affects only one browser and the vendor is unaware, we report it. For widespread security risks affecting all tested browsers, we prioritize reporting the most significant issues due to the volume of risks and the limited time vendors have to review them. Currently, Chrome and Firefox have responded positively to our feedback. Chrome acknowledged a vulnerability we reported involving prompt overlap causing unintended installations. Firefox has responded well to our submissions, fixing one out of the five reported vulnerabilities. Two were deemed harmful, while the rest, though problematic, may not be fixed. We also reported an issue to Samsung Internet regarding navigation to third-party sites without showing the URL origin. Samsung has yet to address this but appreciated our input. We recommend users avoid downloading PWAs with Samsung Internet to reduce phishing attack risks. For more details on the progress with other browsers, refer to our repository [7]. While some vendors address the security risks we report, we can't force them to implement fixes. Therefore, we provide the following tools to assist developers and users.

## 7.3. Developer Perspective

Developers are central to PWA security, yet many introduce flaws unintentionally. GUARDIANPWA-Developer is a local tool that checks manifests for syntactic and semantic issues and recommends corrections, helping developers follow security best practices and protect users.

To implement this, developers place their manifest files into our repository. Our tool can analyze multiple manifests simultaneously, examining all fields for both syntactic and semantic issues. This comprehensive analysis helps prevent errors that could render a PWA unusable, un-installable, or insecure. However, we focus particularly on critical fields such as `name`, `icon`, and `start_url`. For the `name` field, our tool compares the manifest's name with those of all PWAs we have crawled. If there are any similarities, we alert the developer to choose a different name, which can help prevent phishing and brand confusion. Similarly, for the `icon` field, if the icon URL matches that of another PWA, we notify the developer. Even if developers store icons locally and rename them, this check helps mitigate some security risks. Regarding the `start_url`, our tool ensures it points to the correct path, avoiding parent paths and confirming that if it is an HTTPS URL, it is within the scope. Our analysis prevents developers from making mistakes that could compromise the PWA's functionality or security. Additionally, we ensure that developers do not use IDs or other tracking mechanisms that could infringe on user privacy. By using GuardianPWA, developers can improve their PWAs' security, even if they are not familiar with all aspects of PWA installation security.

### 7.4. User Perspective

GUARDIANPWA-User provides a comprehensive solution for assessing the security of PWAs, particularly useful for users on mobile devices who may find it difficult to view or review manifest files. The tool focuses on both the pre-installation and uninstallation phases to ensure maximum security. Below are the five key components of GUARDIANPWA-User:

**Popup Overlaps Detection.** GUARDIANPWA-User detects alerts, prompts, or dialogs overlapping the install banner, which our analysis links to phishing; our experiments confirm detection accuracy. Developer-defined popups, which do not interfere with the banner, are excluded.

**Manifest File Syntax and Security Check.** The tool analyzes the manifest file for syntax errors or fields that might pose security risks. A primary focus is on the `start_url` field, but users are also informed about other critical fields. This is especially important for mobile users who might find it challenging to scrutinize these details. Ensuring the manifest is free from errors and secure fields enhances user safety.

**Service Worker Cache-Only Check.** GUARDIANPWA-User examines the service worker code to ensure it is not configured as cache-only, which could prevent updates. By displaying the service worker code, the tool helps users understand potential issues. This module builds upon and improves existing solutions to provide more reliable detection [51].

**Manifest Update Detection.** The tool tracks changes in the manifest file by storing the initial version upon first access and comparing it with subsequent versions. If a discrepancy is detected, users are immediately notified. This feature ensures users are aware of any updates or changes to the PWA, helping prevent incomplete updates that could compromise security.

**Uninstallation Instructions.** If a PWA has already been installed and users wish to uninstall it, GUARDIANPWA provides clear instructions for safe removal. In cases of uncertainty, users can opt for a straightforward solution: completely uninstalling and reinstalling the browser to ensure all PWA-related data is purged. We provide scripts in our repository for detail instructions [7]. This maintains user security and privacy by eliminating any residual data from potentially insecure PWAs.

### 7.5. Measurement of the Efficiency of the Mitigation Strategies

In our study, we identified a total of 203 security principle violations. These violations can be attributed to two main factors. First, certain browsers continue to violate fundamental security principles, leading to potential privacy leaks. Second, developers may introduce vulnerabilities—either intentionally or unintentionally—through mistakes in code or configuration. If browser vendors strictly adhered to these security principles, even developer errors would not result in privacy leaks.

Currently, only 11 out of the 203 violations have been resolved by browser vendors. To address the remaining issues, we propose two complementary mitigation tools, each operating at different stages of the PWA lifecycle. GUARDIANPWA-Developer focuses on the *pre-installation* phase and helps developers avoid common pitfalls such as misnaming, thereby preventing 81 of the identified violations. On the other hand, GUARDIANPWA-User provides broader protection that covers not only the pre-installation phase but also the installation, post-installation, and uninstallation phases, collectively mitigating up to 150 violations. Although GUARDIANPWA-User can detect many of the same developer-side flaws that GUARDIANPWA-Developer aims to prevent—because both have access to the same manifest and metadata—neither tool addresses the 11 violations that browsers have already fixed. As a result, the comprehensive strategy involves strict browser enforcement combined with both developer-side and user-side tools to effectively reduce or eliminate privacy leaks in PWAs.

## 8. Discussion

**Limitation.** While our study surfaces many security concerns in the PWA installation lifecycle, most do not cause immediate critical threats such as XSS; rather, they violate core security principles and can enable phishing or degrade user experience. Our work has two main limitations. First, our catalog may not capture every browser- and platform-specific vulnerability, especially as vendors introduce new features. Second, GUARDIANPWA depends on adoption by developers and users, does not detect every violation, and cannot address root causes that only browser vendors can resolve internally under the CIA (Confidentiality, Integrity, Availability) principles.

**Differences between PWA and Safari Web Apps.** Unlike PWAs, Safari Web Apps—created via the "Add to Dock" feature in macOS Sonoma—require no Web App Manifest: any website can be docked as a standalone app that runs as an independent process without browser UI elements [52], [53]. This flexibility has a cost: once launched, a Safari Web App does not share session data with Safari, which isolates the app (improving privacy) but breaks seamless cross-context sessions [54], [55].

## 9. Ethical Considerations

Our study followed coordinated disclosure and harm-minimization throughout. All experiments ran on devices, browsers, and PWAs under our own control; the history-sniffing, phishing, and third-party-navigation demonstrations used only researcher-created PWAs and test accounts, never targeting real users or real credentials. Our dataset comprises publicly available PWA manifests crawled from the web; we collected no personal data and conducted no human-subjects study, so no IRB review was required. We reported all identified security-principle violations to the seven affected browser vendors before publication and allowed time to respond: Firefox acknowledged four issues (resolving one and planning two more), Chrome acknowledged the prompt-overlap issue, and Samsung Internet was notified of the origin-hiding navigation issue. To avoid enabling abuse, we release attack demonstrations, tooling, and results [7] only for issues already disclosed to and triaged by the vendors. We believe the benefit of informing vendors, developers, and users outweighs the residual risk of disclosure.

## 10. Related Work

**PWA related work.** PWA is not systematically studied. Several works focus on PWA performance measurement; Steiner evaluates PWAs across different devices and operating systems, identifying the best technologies for hosting them and highlighting Offline Capabilities as the key feature for enabling offline usage [8]. Other related works [56] [57] [58] measure PWA's performance compared to traditional websites and applications in both desktop and mobile devices. Lee et al. [21] initially identified PWA-based attack vectors, detailing side-channel assaults leveraging cache to track PWA usage and potential phishing via push notifications. However, our study shows that many PWAs, such as Starbucks, do not use service workers, prompting a thorough examination of PWAs in practice. Service workers, pivotal for PWA's offline capability, have been pinpointed as security vulnerabilities in numerous studies [59] [60] [61] [21] [62] [63]. Recent work by Wang et al. systematically demystifies the PWA permission system, revealing how permission inconsistencies across browsers and platforms can be exploited to mislead users and escalate privileges [64]. These vulnerabilities could enable severe attacks, including botnets, phishing, and Library-Hijack.

**Related Work on Android Installation Security Issues.** Research on Android installation security has highlighted several key areas of concern, including permission management, malware detection, and secure installation processes. Effective permission models and user awareness are essential to mitigate risks posed by apps requesting excessive permissions [65]. Advanced malware detection techniques, particularly those leveraging machine learning, have been developed to identify and prevent malicious applications during the installation process [66]. Ensuring secure installation involves verifying app authenticity and utilizing cryptographic techniques to prevent tampering and interception [67].

**Related Work on Browser Extension Manifest Security.** In the realm of browser extension security, Manifest V3 has introduced significant changes aimed at enhancing extension security, privacy, and performance. These changes include replacing background scripts with service workers, enforcing stricter Content Security Policies (CSP), and limiting permissions that extensions can request [68]. Research has categorized various types of malicious behaviors in extensions, such as data theft and click fraud, emphasizing the need for robust detection mechanisms [69]. A combination of dynamic analysis and manual code reviews is often employed to uncover sophisticated malicious behaviors that automated tools might miss [70].

## 11. Conclusion

Our analysis of the PWA installation lifecycle reveals critical security risks, demonstrating that failing to adhere to the CIA (Confidentiality, Integrity, Availability) principles can lead to significant privacy leaks. We identified 203 security violations in current PWA implementations and found that many PWAs lack service workers, challenging existing definitions of the PWA lifecycle. Our findings show risks in manifest fields like `name`, `icon`, and `start_url`, such as duplication, misleading icons, and improper paths. The GUARDIANPWA framework helps developers address these vulnerabilities, but our work also proves that neither developers nor users can fully trust browser vendors. As new PWA installation features emerge, we urge vendors to strictly follow the CIA principles. By addressing security risks and providing actionable recommendations, GUARDIANPWA enhances PWA security and ensures a more trustworthy installation process.

## LLM Usage Statement

We used a large language model to assist with language editing, LaTeX formatting, and condensing prose. All research design, experiments, analysis, and conclusions are the authors' own, and the authors verified all content.

# Appendix A. Comprehensive Catalog of PWA Security Risks

TABLE 6: Comprehensive Catalog of PWA Security Risks

| Security Risk | CIA | Browser Count |
|---|---|---|
| Inconsistent profile leads to user confusion | C | 3 |
| Private mode PWA allowed and used in normal mode | C | 1 |
| Discrepancies in PWA installation requirements | I | 16 |
| Name and icon duplication causing user confusion and phishing | C | 16 |
| ID can be duplicated and multiple PWAs not distinguishable | I | 16 |
| `start_url` and scope leading to external sites | I | 16 |
| Icon can be a third-party URL, leading to phishing | C | 16 |
| Display fullscreen mode hides URLs, enabling phishing attacks | C | 16 |
| Related applications can lead to third-party app installations | C | 1 |
| Manifest allows arbitrary fields, leading to tracking | I | 16 |
| History sniffing via Chrome Mini-Infobar | C | 1 |
| Alert, Confirm, Prompt banners overlap with installation banner | C | 6 |
| Lack of confirmation steps leading to accidental installations | I | 7 |
| Automatic prompt installation banners | A | 2 |
| Redundant installation prompts after PWA installation | A | 1 |
| Lack of visibility of PWA's origin | C | 1 |
| Allowing multiple identical PWAs increases attack surface | I | 10 |
| No PWA manifest update | I | 10 |
| Inconsistent PWA manifest update detection | I | 7 |
| Navigation to third-party scopes without awareness | C | 1 |
| Service worker cache-only strategies prevent updates | I | 16 |
| Complexity in PWA uninstallation process | C | 16 |
| Uninstalled PWAs can still send notifications | A | 2 |
| Chrome's PWA management causing confusion across profiles | C | 3 |
| Difficulties in managing PWAs across profiles in Edge | C | 3 |
| **Total** | | **203** |

## A.1. Experimental Dataset and Setup

We conducted our validation on popular browsers with over 1% market share, focusing on their support for Progressive Web App (PWA) installations. Table 7 lists the specific versions tested. Since Chrome, Edge, and Firefox undergo frequent updates that can alter PWA behaviors, we examined versions 117 through 120 as well as the latest iteration (130), allowing us to track incremental changes in security and installation flows. Safari, tested in versions 17 and 18, represents the continual evolution of PWA functionality on iOS. For other Chromium-based browsers such as Samsung Internet, Opera, and Brave, we concentrated on their most recent stable releases (22, 77, and 1.58, respectively). Although these browsers typically follow Chromium's core policies, we found minor delays and divergences in how they adopt new PWA features.

TABLE 7: Browsers and Versions Used for PWA Testing

| Browser | Versions Tested |
|---|---|
| Chrome | 117 – 120, 130 |
| Edge | 117 – 120, 130 |
| Firefox | 117 – 120, 130 |
| Safari | 17, 18 |
| Samsung Internet | 22 |
| Opera | 77 |
| Brave | 1.58 |

Our experiments spanned multiple environments to capture diverse installation scenarios. On Android, we employed both a BLU G53 64GB device and an Android Studio emulator for reproducibility. We tested iOS browsers (including Safari) on devices running iOS 17 and 18. Desktop experiments were performed on macOS (using an Apple MacBook Pro M3) and on Windows and Linux via virtual machines. An additional HP laptop equipped with an Intel Core i7 processor and Microsoft Store access was used to verify any installation dependencies specific to Windows Store apps.

The dataset originated from the raw collection described in [71], which relies on Common Crawl. However, since that source adopted a broad definition of PWAs and inadvertently included non-PWA sites, we refined our sample by requiring valid manifest files containing essential fields (`name`, `display`, etc.) critical for the installation process. We also checked for the presence of a service worker, while recognizing that a missing service worker does not outright invalidate a web application's PWA status. This stricter filtering yielded a more accurate dataset, ensuring that our tests addressed legitimate PWAs rather than general websites with incomplete or misleading metadata.

## A.2. Extended Semantic and Syntactic Analysis of Web App Manifest Fields

In this section, we will cover *id*, *display*, and *related_applications*. More detailed results for all fields are available on our website [7].

**ID.** The *id* field is designed to uniquely identify a PWA. However, during fuzzing measurements, it was observed that the *id* field can differ for the same origin, leading to multiple PWAs having the same unique App ID, which violates the definition of unique identity by major browser vendors and the W3C [72], [73]. This inconsistency can cause issues in uniquely identifying applications across different origins.

The current algorithm for updating the manifest ID based on JSON input lacks a check to ensure the *id* is unique within the same origin. According to the W3C and MDN, the *id* should ensure unique identity even if the web application is from the same URL [74]. Based on our analysis of our dataset, PWAs use IDs in different ways. Some of them use their own or third-party URLs as their IDs. The W3C defines the best practice as using `/` as their IDs because developers should not use this field to track users.

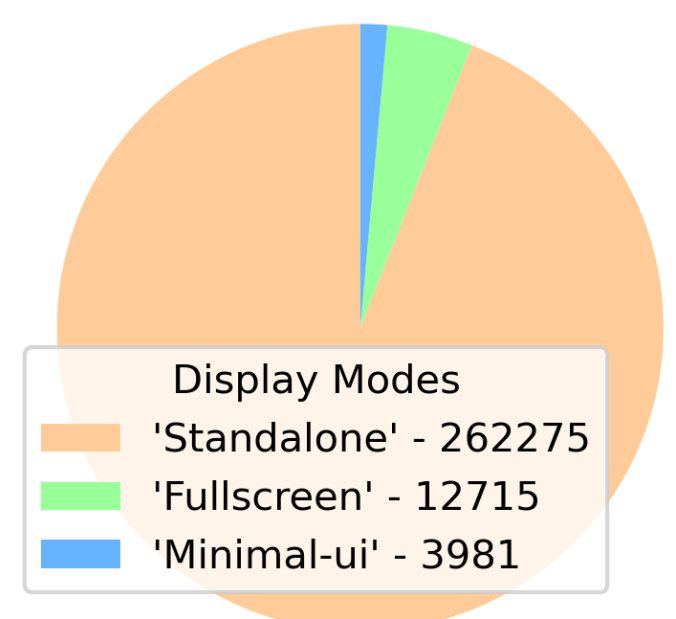

Figure 11: Distribution of Display Modes

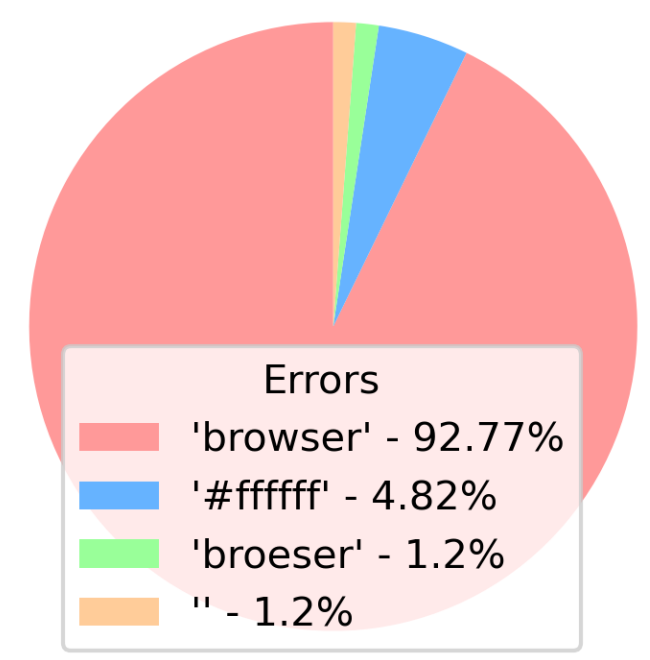

Figure 12: Error Rates in Display Modes

**Display.** The *display* field controls the appearance of the application. Values include:

- *browser*: The PWA opens as a tab in the browser.
- *minimal-ui*: Best for practical use, as it is fullscreen but shows the URL.
- *fullscreen*: Does not display time or other UI elements, only the content.
- *standalone*: Resembles a native app view, showing the time and back button.

Unknown values will prevent the PWA from being installable. To be installable, the *display* property must be one of *standalone*, *fullscreen*, or *minimal-ui*. Common display errors are shown in Figure 12. In Figure 11, although many developers use *standalone*, *minimal-ui* is considered the best practice.

**Related_applications and prefer_related_applications.** These fields are interrelated. Having only one of them is not useful. These fields may change in the future as they are experimental features supported only on Android Chrome. They raise security concerns: if a user clicks an install button created by developers, the app is downloaded from the Google Play Store; otherwise, it installs the PWA manually. The *id* is more important than the URL in this context. Specifically, users might see a button linking to an official app, but the PWA itself could be a phishing attempt. Users may trust the PWA, leading to potential privacy breaches.

## A.3. Installation Banner Analysis

**Overlapped Prompts Analysis.** Other overlapping prompts, such as popovers and popup technology, do not obstruct the PWA install banner. Instead, the banner takes precedence and will overlay the popover. This ensures that even if a popover is malicious, it will not interfere with the installation process. Therefore, popover technology poses no issues in this context.
**Crafted Banners.** Users might think they installed PWA1, but they actually installed PWA2. This issue arises due to misleading banners and is particularly problematic in Chrome on Android, where related applications can lead users to native apps instead of the intended PWA. Additionally, leading to a PWA via the *start_url* or a redirect can be problematic. Samsung Internet is especially susceptible, as it does not display the URL, making it easier for users to be deceived.

## A.4. Detailed Categorization of PWA Security Risks

Table 6 presents a comprehensive summary of the security risks associated with PWAs. Each identified risk is classified according to its impact on the core security principles—Confidentiality (C), Integrity (I), and Availability (A). The table also highlights the number of browsers affected by each issue, emphasizing the widespread and critical nature of these vulnerabilities.